\documentclass[12pt]{iopart}
\usepackage{adjustbox}
\usepackage[utf8]{inputenc}
\usepackage{graphicx}
\usepackage{dcolumn}
\usepackage{bm}
\usepackage{placeins}
\usepackage{array}
\newcolumntype{L}[1]{>{\raggedright\arraybackslash}p{#1}}
\newcolumntype{C}[1]{>{\centering\arraybackslash}p{#1}}
\newcolumntype{R}[1]{>{\raggedleft\arraybackslash}p{#1}}

\begin{document}

\newcommand{\gguide}{{\it Preparing graphics for IOP Publishing journals}}

\title[J. Phys.: Condens. Matter]{Tight-binding model of Pt-based jacutingaites as combination of the honeycomb and kagome lattices}

\author{G. Santos-Castro \footnote{Author to whom any correspondence should be addressed.}} 
\address{Grupo de Teoria da Matéria Condensada, Departamento de Física, Universidade Federal do Ceará, 60455-760 Fortaleza, Ceará, Brazil}
\ead{giselle@fisica.ufc.br}
\vspace{10pt}

\author{L. K. Teles} 
\address{Grupo de Materiais Semicondutores e Nanotecnologia, Instituto Tecnológico de Aeronáutica, DCTA, 12228-900 São José dos Campos, Brazil}
\ead{lkteles@ita.br}
\vspace{10pt}

\author{I. Guilhon Mitoso} 
\address{Grupo de Materiais Semicondutores e Nanotecnologia, Instituto Tecnológico de Aeronáutica, DCTA, 12228-900 São José dos Campos, Brazil}
\ead{guilhon@ita.br}
\vspace{10pt}

\author{J. M. Pereira Jr.} 
\address{Grupo de Teoria da Matéria Condensada, Departamento de Física, Universidade Federal do Ceará, 60455-760 Fortaleza, Ceará, Brazil}
\ead{pereira@fisica.ufc.br}
\vspace{10pt}

\begin{indented}
\item[]September 2024
\end{indented}

\begin{abstract}
We introduce a refined tight-binding (TB) model for Pt-based jacutingaite materials Pt$_{2}N$X$_{3}$, ($N$ = Zn, Cd, Hg; X = S, Se, Te), offering a detailed representation of the low-energy physics of its monolayers. This model incorporates all elements with significant spin-orbit coupling contributions, which are essential for understanding the topological energy gaps in these materials. Through comparison with first-principles calculations, we meticulously fitted the TB parameters, ensuring an accurate depiction of the energy bands near the Fermi level. Our model reveals the intricate interplay between the Pt $3e$ and $N$ metal orbitals, forming distinct kagome and honeycomb lattice structures. Applying this model, we explore the edge states of Pt-based jacutingaite monolayer nanoribbons, highlighting the sensitivity of the topological edge states dispersion bands to the nanostructures geometric configurations. These insights not only deepen our understanding of jacutingaite materials but also assist in tailoring their electronic properties for future applications.

\end{abstract}

\submitto{\JPCM}

\maketitle

\section{Introduction}

The successful synthesis of graphene \cite{novoselov2004electric}, coupled with the discovery of its remarkable properties, has spurred extensive research on the properties and unique characteristics of novel two-dimensional (2D) materials \cite{katsnelson2007graphene, neto2009electronic, dong2010properties, bhimanapati2015recent, ares2021recent}. The investigation of alternative 2D materials, such as $X$enes (X = Si, Ge, Sn), transition metal dichalcogenides, and black phosphorus, has revealed a variety of unique properties and energy gaps, providing superior control over the electronic behavior and transport \cite{xu2013graphene, balendhran2015elemental, gupta2015recent, gao2017flexible, guo20192d, glavin2020emerging}. Moreover, materials exhibiting intrinsic spin-orbit coupling (SOC) energy gaps leading to the Quantum spin Hall (QSH) effect have drawn great attention. This effect is characterized by a Z$_2$ topological invariant distinguishing these materials from ordinary insulators. The emergence of such topological states implies a paradigm shift with implications for spintronics \cite{hasan2010colloquium, isaeva2013bismuth, bansil2016colloquium, olsen2019discovering, marrazzo2019relative, cayssol2021topological}. Inherently resistant to backscattering from various perturbations, these states pave the way for robust, non-dissipative charge transport and innovative applications in quantum computing and photonics \cite{hasan2010colloquium, kong2011opportunities, bhardwaj2020topological}.
The concept of QSH states, initially theorized for graphene by Kane and Mele \cite{kane2005z}, found empirical evidence in HgTe/CdTe and InAs/GaSb quantum wells \cite{bernevig2006quantum, Knez2011}. These states, protected by time-reversal symmetry, are robust against non-magnetic perturbations. A notable realization of such topological systems is the jacutingaite (Pt$_{2}$HgSe$_{3}$, Z$=2$) monolayer \cite{kandrai2020signature}, approximated as a honeycomb structure of Hg elements.

The jacutingaite is a new species of platinum-group mineral discovered at the Cauê iron-ore deposit, Itabira district,
Minas Gerais, Brazil, in 2008 \cite{cabral2008platinum}, and then synthesized in 2012 \cite{vymazalova2012jacutingaite}. The ore is brittle, highly stable, with an excellent cleavage $\{$001$\}$, whose perpendicular direction is composed of layers stacked by van der Waals interaction in an AA configuration \cite{vymazalova2012jacutingaite}.
In its bulk form, the jacutingaite is a hexagonal structure that exhibits dual-topological semi-metal characteristics, combining a weak QSH phase and mirror-protected topological crystalline phase \cite{facio2019dual, cucchi2020bulk}. In its monolayer form manifests as a QSH topological semiconductor \cite{marrazzo2018prediction, kandrai2020signature}, as evidenced by a measured topological gap of 110~meV and decay of edge states into the bulk over approximately 5 \AA~\cite{kandrai2020signature}. The jacutingaite monolayer is the most promising material to host QSH states similarly to the original proposal for graphene described by Kane-Mele model, given its structural stability and large topological gap \cite{marrazzo2018prediction, kandrai2020signature, bhardwaj2020topological, weber20242024} .

The jacutingaite monolayer family $M_{2}NX_{3}$ ($M=$ Ni, Pd, Pt; $N=$ Zn, Cd, Hg; $X=$ S, Se, Te) were theoretically presented by de~LIMA, F. et al., 2020 \cite{de2020jacutingaite}. They demonstrated that there are other jacutingaite like structures that also host the Kane-Mele QSH state. Relevant for top-down mode synthesis, they also demonstrated that the bulk structure presents cleavage energies at the level of other experimentally exfoliable materials, in the range of 0.22 J/m$^{2}$ -- 0.85 J/m$^{2}$ \cite{de2020jacutingaite}. From this family, one group that we can highlight are those that present a energy gap and Kane-Mele QSH state, the Pt-based jacutingaites.

Distinct from graphene, the dispersion is only approximately linear from the K(K$'$)--$\Gamma$ directions, whereas along the K(K$'$)--M direction it is almost flat.  The traditional Kane-Mele model inadequately captures these low-energy characteristics of jacutingaite monolayers, requiring a more nuanced approach for accurate representation.

This work investigates the Pt-based jacutingaite monolayers (Pt$_{2}NX_{3}$, N= Zn, Cd, Hg; X= S, Se, Te) and introduces an advanced tight-binding (TB) model derived from density functional theory (DFT) band structures. Predicted as QSH topological semiconductors with substantial SOC band gaps \cite{de2020jacutingaite}, these monolayers are modeled considering all SOC-contributing elements and their respective lattices, comprising a honeycomb and a kagome lattice superposition. Furthermore, we incorporate real next-nearest-neighbor (NNN) hopping terms and all hopping terms within a radius equivalent to the NNN distance to the N sites, which form the basis of the Kane-Mele model. Our refined TB model exhibits excellent agreement with first-principles calculations across a broad spectrum of wavevectors and energies around the Fermi level, promising enhanced understanding of these fascinating materials.

\section{Methodology}

\subsection{Tight Binding Model for jacutingaite-like
systems}

\begin{figure}[h]
\includegraphics[width=9cm]{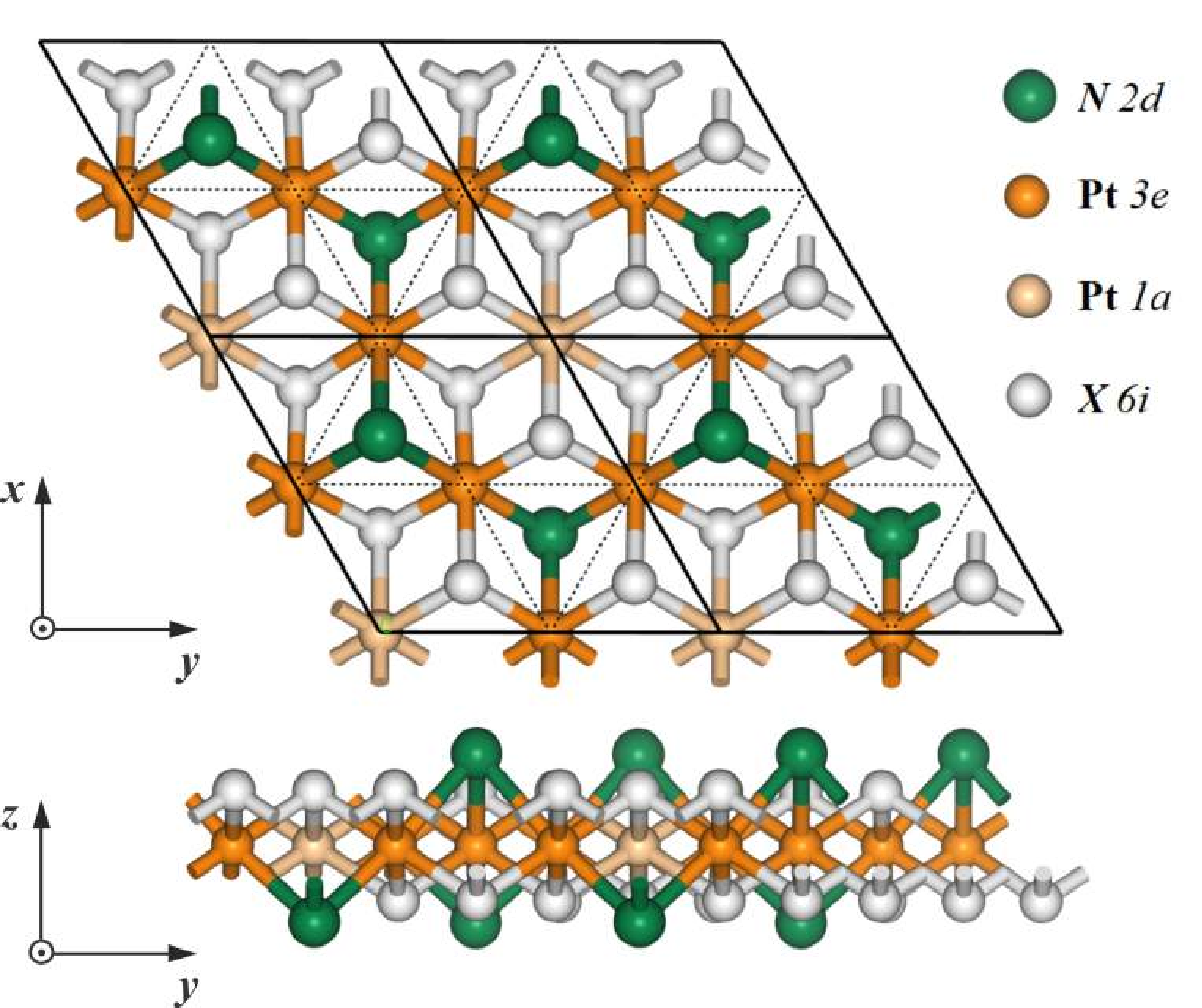} 
\caption{Top and profile view of Pt$_{2}NX_{3}$ monolayer. Green and gray balls represent $N$ and $X$ elements, respectively. Orange and yellow balls, Pt $3e$ and Pt $1e$, conform Wyckoff position, respectively. Inside black line area represents unit cell. The dotted lines connect Pt $3e$ elements.} \label{structurefull}
\end{figure}

The crystal structure of jacutingaite ($M_{2}NX_{3}$, $\textrm{Z}=2$) has space group $P\bar{3}m1$ \cite{cabral2008platinum, vymazalova2012jacutingaite}. The Pt-based jacutingaite monolayer and Wyckoff position of their elements are illustrated in  Figure~\ref{structurefull}, where $M=\mathrm{Pt}$. The electronic characteristics of jacutingaite-family materials can be understood by taking into account their structure, which, with respect to the low-lying bands can be described as two interpenetrating honeycomb and kagome sub-lattices.

Jacutingaite-type systems have structural and electronic characteristics of both types of lattices. Thus, we propose a TB model that combines these two sub-lattices, i.e., a honeycomb-kagome (HK) structure, in order to obtain a better agreement with the DFT energy bands near the SOC energy gap, in comparison with the pure Kane-Mele model.

\begin{figure}[h]
\includegraphics[width=9cm]{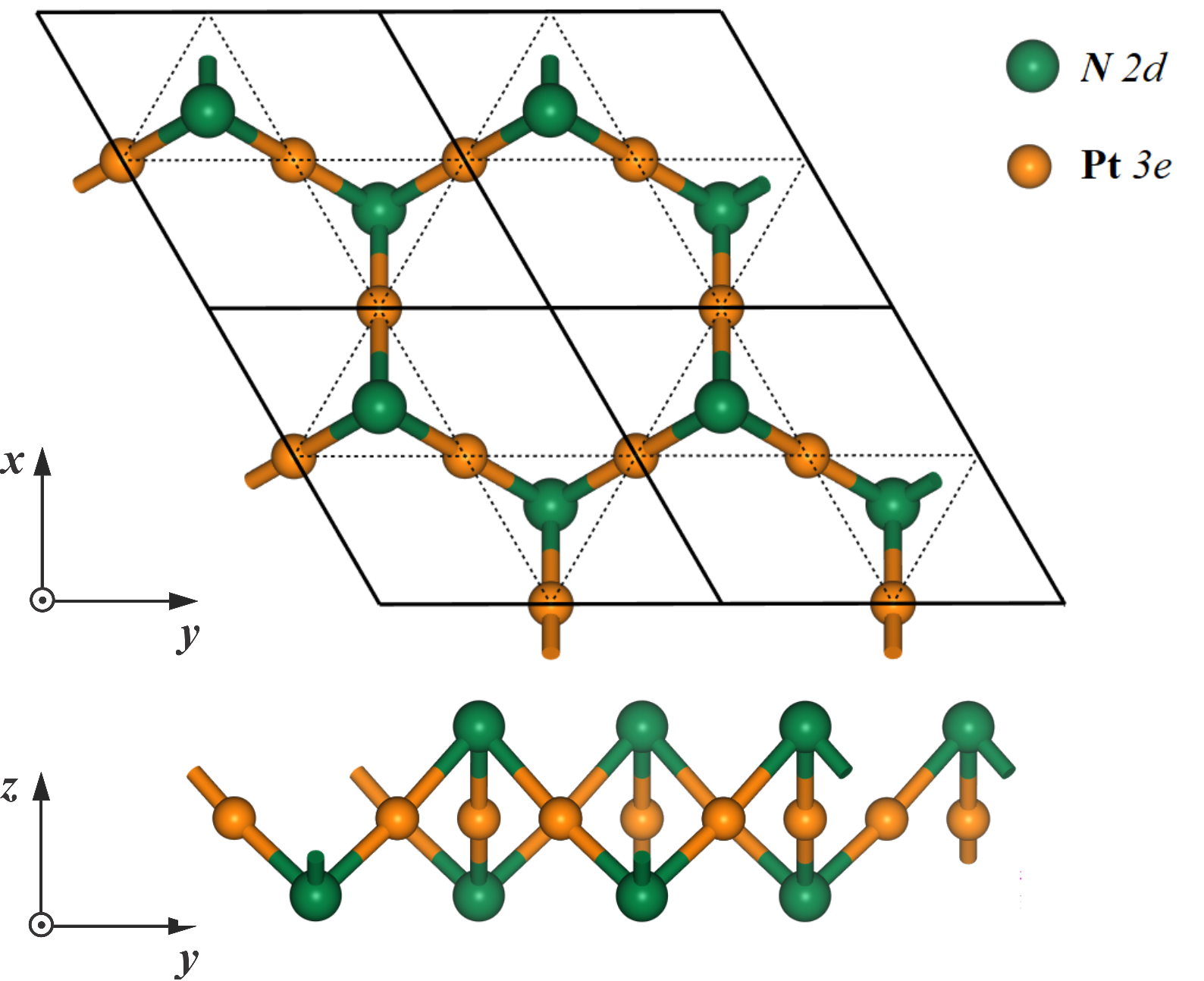}
\caption{Top and profile view of the geometric structure considered to development of the TB model for jacutingaite-like systems. Green and orange spheres represent $N$ and Pt $3e$ elements, respectively. Inside black line area represents unit cell. The dotted lines connect Pt $3e$ elements form kagome planar lattice and $N$ metals intermediated by Pt $3e$ elements form the honeycomb buckled lattice. The system then consists of two interpenetrating honeycomb and kagome sub-lattices.} \label{structuresimple}
\end{figure}

As shown in the next section the orbitals that contribute most significantly to the conduction and valence bands around the topological energy gap are S$_{1/2}$ orbitals from $N$ metal e 5D$_{5/2}$ orbitals from Pt $3e$. By taking into account only these elements, one can visualize (see Figure \ref{structuresimple}) the jacutingaite-like structure as a honeycomb buckled lattice, resembling the silicene or germanene structure, composed of $N$ metals and a planar kagome lattice formed by the Pt $3e$ elements.

\begin{figure}[h]
\includegraphics[width=9cm]{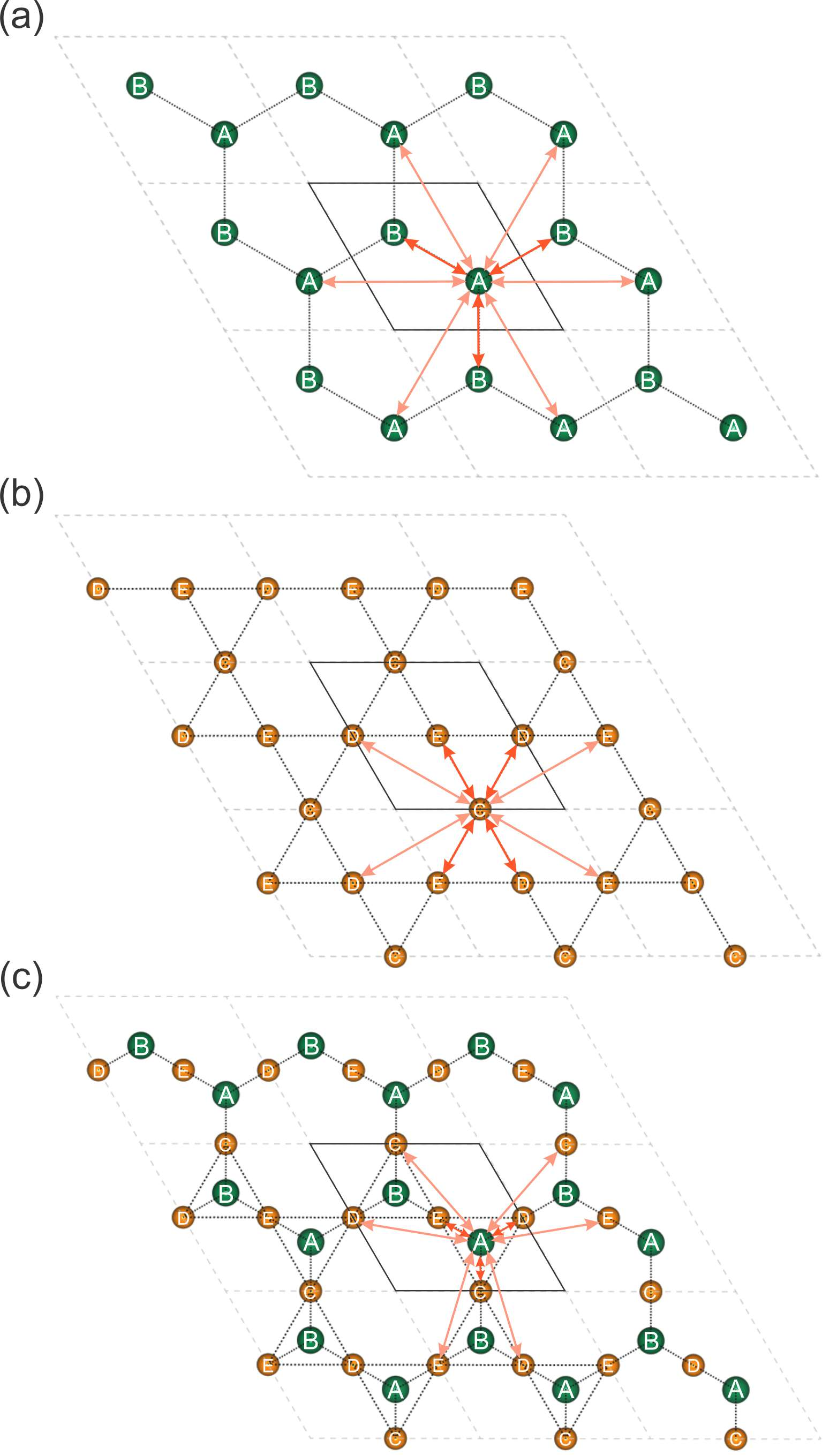}
\caption{Top view of: (a) honeycomb buckled lattice formed by $N$ metals, in dotted lines. This structure has two nonequivalent sub-lattices, $A$ and $B$. (b) Kagome planar lattice composed of Pt element, in dotted lines. In this structure there are three nonequivalent sub-lattices, $C$, $D$ and $E$.  (c) 
Two interpenetrating honeycomb buckled and kagome planar sub-lattices composing the basis of the tight binding model for jacutingaite-like systems. The dark orange arrows are NN hoppings and light orange arrows are NNN hoppings between (a) $N$$-$$N$, (b) Pt$ - $Pt and (c) Pt$ - N$. For all cases: inside gray dashed area, the unit cells; inside black line area, home-cell.} \label{structurehopping}
\end{figure}

In Figure \ref{structurehopping} (a) the dotted lines connecting $N$ metals form the hexagons of a honeycomb lattice. The region highlighted by the black solid contour depicts a unit cell, showing that two $N$ metals per cell occupy non-equivalent sites, which are labeled as $A$ and $B$. We considered the on-site energy, nearest-neighbor (NN) and NNN hopping parameters as indicated by the arrows, as well as the SOC for the construction of the corresponding Hamiltonian term based on the buckled sub-lattice formed by $N$ metals.

In  Figure \ref{structurehopping} (b) the dotted lines connecting the Pt elements form a triangular geometry with an arrangement of interconnected hexagons. There are three Pt elements per cell occupying non-equivalent sites, which are labeled as $C$, $D$ and $E$. As in the previous case, we included an on-site energy, NN and NNN hopping parameters and SOC, as indicated by the arrows, for the construction of the Hamiltonian term for the planar sub-lattice formed by Pt elements.

\begin{figure}[h]
\includegraphics[width=8cm]{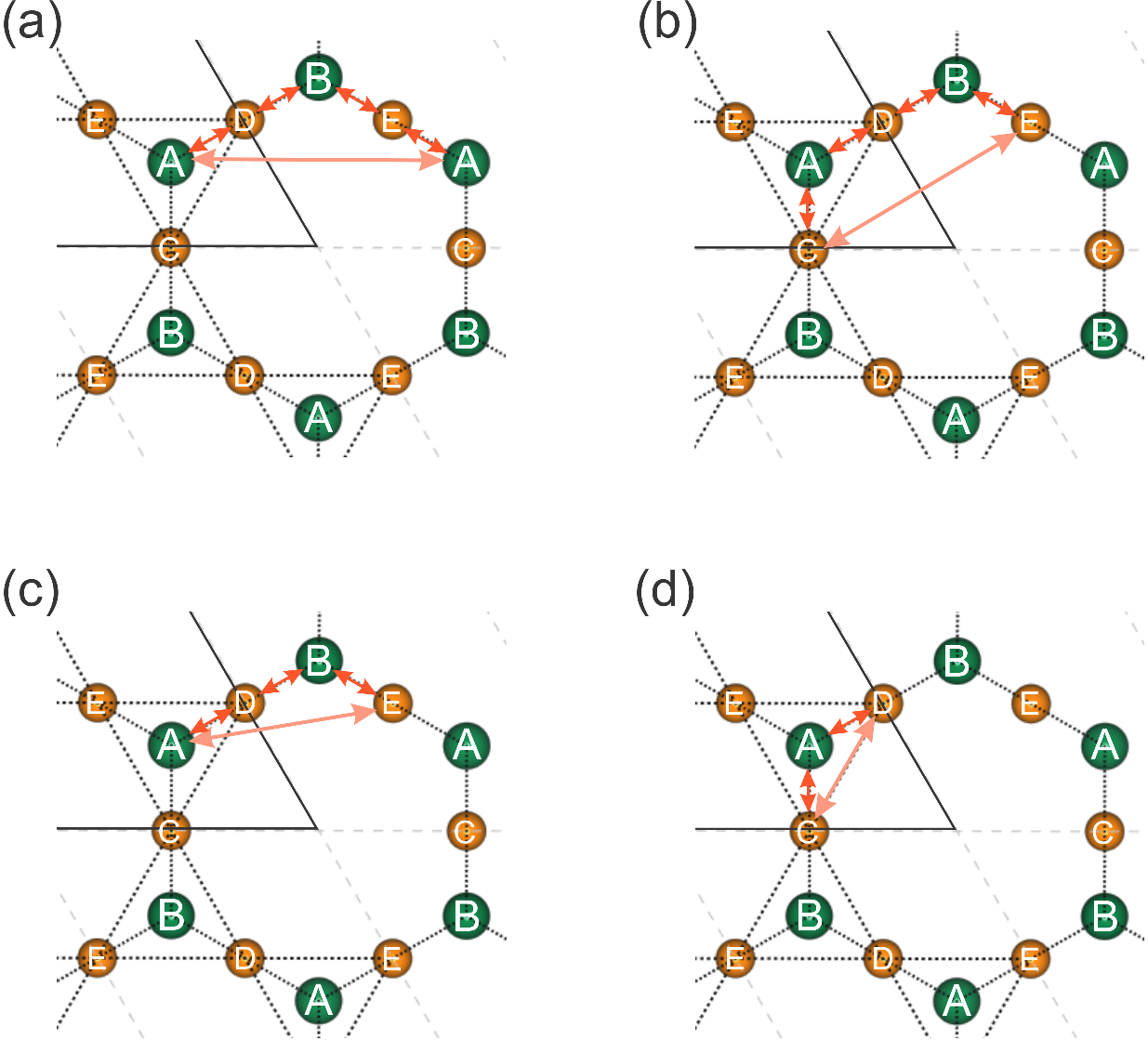}
\caption{Illustration of vectors connecting neighboring sites used in the term SOC (a) $N$-NNN in a honeycomb lattice (b) Pt-NNN in a kagome lattice (c) Pt-$N$ NNN in HK (d) Pt--Pt NN in HK.} \label{struct_soc}
\end{figure}

Figure \ref{structurehopping} (c) depicts the system as two interpenetrating honeycomb buckled and kagome planar sub-lattices. Thus, we considered the NN and NNN hopping parameters as indicated by the arrows, connecting a Pt to a $N$ metal, and two SOC terms, namely Pt--$N$--Pt and Pt--$N$--Pt--$N$--Pt. Figure \ref{struct_soc} displays all cases of SOC mentioned. All possible hoppings for the distances smaller than that between NNN $N$ metal are used in this model.

We define the Hamiltonian for jacutingaite-like systems as:
\begin{equation}\label{hhk}
\hat{H}_{HK}=\hat{H}_{\mathrm{h}}+\hat{H}_{\mathrm{k}} + \hat{H}_{\mathrm{hk}}
\end{equation}
where the honeycomb and kagome sub-lattices terms are:
\begin{equation}
\hat{H}_{\mathrm{h}}, \hat{H}_{\mathrm{k}}=\hat{H}_{\mathrm{on-site}}+\hat{H}_{\mathrm{NN}}+\hat{H}_{\mathrm{NNN}}+\hat{H}_{\mathrm{SOC}},
\end{equation}
with
\begin{equation}
\hat{H}_{\mathrm{on-site}}=\sum_{p, \alpha}\epsilon_{p, \alpha}c^{\dagger}_{p; \alpha}c_{p, \alpha},
\end{equation}
\begin{equation}
\hat{H}_{\mathrm{NN}}=-\sum_{\langle p q \rangle;\alpha}t^{'}_{pq}c^{\dagger}_{p, \alpha}c_{q, \alpha},
\end{equation}
\begin{equation}
\hat{H}_{\mathrm{NNN}}=-\sum_{\langle\langle p q \rangle\rangle;\alpha}t^{''}_{pq}c^{\dagger}_{p, \alpha}c_{q, \alpha},
\end{equation}
\begin{equation}
\hat{H}_{\mathrm{SOC}}=i\sum_{\langle\langle p q \rangle\rangle;\alpha\beta}\lambda^{''}_{pq}c^{\dagger}_{p, \alpha}\big(\nu_{pq}\cdot\bm{\sigma}\big) c_{q, \beta}.
\end{equation}
The first term contains the on-site energies given by $\epsilon_{p, \alpha}$, where $p$ and $\alpha$ represent an orbital and spin, respectively; the $\hat{H}_{\mathrm{NN}}$ and $\hat{H}_{\mathrm{NNN}}$ terms describe NN and NNN hopping between $p$ and $q$ orbitals, being $t'_{pq}$ and $t''_{pq}$ the respective parameters and $\hat{H}_{\mathrm{SOC}}$ the SOC terms, with $\lambda^{''}_{pq}$ as SOC parameter between $p$ and $q$ orbitals and also between $\alpha$ and $\beta$ spin, where
$c^{\dagger}_{p,\alpha}$($c_{p,\alpha}$)
is the fermion creation (annihilation) operator at site $p$ and with spin $\alpha$ and $\nu_{pq}$ is a unit vector that connects the sites of localized $p$ and $q$ orbitals by a single intermediate site. Finally, $\bm{\sigma} = (\sigma_{x}, \sigma_{y}, \sigma_{z})$ is the vector of Pauli matrices. Figure \ref{structurehopping} (c) illustrates the terms above.

For the honeycomb sub-lattice we have $\nu_{pq}$ as:
\begin{equation}\label{soc_h}
\nu_{im}=\Bigg[\frac{\big(\bm{d}_{ij}+\bm{d}_{jk}\big)\times\big(\bm{d}_{kl}+\bm{d}_{lm}\big)}{ | \big(\bm{d}_{ij}+\bm{d}_{jk}\big)\times\big(\bm{d}_{kl}+\bm{d}_{lm}\big)|}\Bigg], 
\end{equation}
whereas for the kagome sub-lattice it is given by:
\begin{equation}\label{soc_k} 
\nu_{jn}=\Bigg[\frac{\bm{d}_{jk}\times\bm{d}_{kl}+\bm{d}_{lm}\times\bm{d}_{mn}}{ | \bm{d}_{jk}\times\bm{d}_{kl}+\bm{d}_{lm}\times\bm{d}_{mn} | }\Bigg],
\end{equation}
with the $\{i,k,m\}$ indices refer to $h$S$_{1/2}$ orbitals from different sites belonging to the honeycomb sublattice, where $h=4,5 \textrm{ and } 6$ for $N=$ Zn, Cd and Hg, respectively, whereas the $\{j,l,n\}$ indices representing 5D$_{5/2}$ orbitals from different sites belonging to the kagome sublattice. Figures \ref{struct_soc} (a) and (b) illustrates the definitions of Eqs. \ref{soc_h} and \ref{soc_k}, respectively.

Next, $\hat{H}_{\mathrm{hk}}$ is the term that couples elements from different sub-lattices:
\begin{equation}\label{keyEq}
 \hat{H}_{\mathrm{hk}} =\hat{H}_{\mathrm{NN}}^{ij}+\hat{H}_{\mathrm{NNN}}^{ij}+\hat{H}_{\mathrm{SOC}}^{jl}+ \hat{H}_{\mathrm{SOC}}^{il}.
\end{equation}
The $\hat{H}_{\mathrm{NN}}^{ij}$ and $\hat{H}_{\mathrm{NNN}}^{ij}$ are Hamiltonian describe NN and NNN hopping between $i$ and $j$ orbitals, being $t'_{ij}$ and $t''_{ij}$ the respective parameters:
\begin{equation}
\hat{H}_{\mathrm{NN}}^{ij}=-\sum_{\langle ij\rangle;\alpha}t^{'}_{ij}c^{\dagger}_{i, \alpha}c_{j, \alpha}
\end{equation}
\begin{equation}
\hat{H}_{\mathrm{NNN}}^{ij}=-\sum_{\langle\langle i j \rangle\rangle;\alpha}t^{''}_{ij}c^{\dagger}_{i, \alpha}c_{j, \alpha}
\end{equation}
and $\hat{H}_{\mathrm{SOC}}^{jl}$ as SOC Hamiltonian, with $\lambda^{''}_{jl}$ as SOC parameter between $j$ and $l$ orbitals of Pt--$N$--Pt:
\begin{equation}
\hat{H}_{\mathrm{SOC}}^{jl}=i\sum_{\langle jl\rangle}\lambda^{'}_{jl}c^{\dagger}_{j, \alpha}\big(\nu_{jl}\cdot\bm{\sigma}\big) c_{l, \beta}
\end{equation}
being:
\begin{equation}\label{4d}
\nu_{jl}=\Bigg[\frac{\bm{d}_{jk}\times\bm{d}_{kl}}{ | \bm{d}_{jk}\times\bm{d}_{kl} |}\Bigg]
\end{equation}
and $\hat{H}_{\mathrm{SOC}}^{il}$ as SOC Hamiltonian, with $\lambda^{''}_{il}$ as SOC parameter between $i$ and $l$ orbitals of Pt--N--Pt--N--Pt:
\begin{equation}
\hat{H}_{\mathrm{SOC}}^{il}=i\sum_{\langle\langle i l\rangle\rangle}\lambda^{''}_{il}c^{\dagger}_{i, \alpha}\big(\nu_{il}\cdot\bm{\sigma}\big) c_{l, \beta}
\end{equation}
being:
\begin{equation}\label{4c}
\nu_{il}=\Bigg[\frac{(\bm{d}_{ij}+\bm{d}_{jk})\times\bm{d}_{kl}}{ | (\bm{d}_{ij}+\bm{d}_{jk})\times\bm{d}_{kl} |}\Bigg]
\end{equation}
Figures \ref{struct_soc} (d) and  (c) illustrates the definitions of Eqs. \ref{4d} and \ref{4c}, respectively.

Thus, the present TB model has twelve independent parameters, which can be adjusted using as reference the results from ab initio calculations, as shown below.

\subsection{Ab initio calculations}

In order to investigate the structural and electronic properties of the Pt-based jacutingaite monolayers, first-principles DFT-based calculations \cite{hohenberg1964inhomogeneous, kohn1965self} were performed using the plane waves method as implemented on the Quantum ESPRESSO software package  \cite{giannozzi2009quantum, giannozzi2017advanced}.  A vacuum distance of 20 \AA~and dipole correction were inserted to decouple the periodic images  \cite{sohier2017density}.

The generalized gradient approximation of the Perdew–Burke–Ernzerhof (GGA-PBE) functional  \cite{perdew1996generalized} was used to describe the exchange-correlation interaction. In order to solve Kohn-Sham equations, the valence electrons were treated explicitly while central electrons have been replaced by the optimized norm-conserving Vanderbilt pseudopotential  \cite{hamann2013optimized, van2018pseudodojo}. 

The structural optimization was performed by using Broyden–Fletcher–Goldfarb– Shanno interactive (BFGS) to minimize the total energy calculated by DFT 
\cite{fletcher1987practical, billeter2000linear, billeter2003efficient}, with energy threshold of $4.0\times 10^{-5}$~Ry and Hellmann-Feynman forces on each atom smaller than  $4.0\times 10^{-5}$~Ry/\AA.
The self-consistent energy threshold was of $1.0\times 10^{-8}$~Ry, the plane-wave energy cutoff was set as 120 Ry and the Brillouin zone (BZ) is sampled by a $\Gamma$-centered $12\times12\times1$ Monkhrost-Pack grid. 

Calculations were performed both with and without SOC. The calculation without the consideration of SOC were used as an intermediate step for the optimization of atomic structure and TB hopping parameters. The definite results are determined from the DFT calculations with the inclusion SOC considering the previously described results as starting points.

\subsection{Optimization of hopping parameters}
The PythTB package \cite{yusufalytight} was used for the TB calculations. The parameters of on-site energies and hopping amplitudes were adjusted to fit the DFT band structure in part of the BZ, for case without and with SOC, for applying the minimization process using the BFGS algorithm \cite{fletcher1987practical}, for each $\vec{k}$ along the line between $\textrm{K}$($\textrm{K}'$) and $\textrm{M}$ points, where the cost function to be minimized
\begin{equation} \mathrm{J}=\sum_{n, k}\Delta\epsilon_{n,\textbf{k}}=\sum_{n, k}\Big(\epsilon^{DFT}_{n,\textbf{k}}-\epsilon_{n,\textbf{k}}^{TB}\Big)^{2},
\end{equation}
with $\epsilon^{DFT}_{n,\textbf{k}}$ and $\epsilon^{TB}_{n,\textbf{k}}$ being the energy of the $n$ band in momentum space as given by DFT and TB approaches, respectively. The index $n$ may assume four values, including the top of the valance band and the bottom of conduction band. The window of momentum space is such that $\textbf{k} \in$ $g$--M--K--$g'$, where $g$ and  $g'$ are points in $\Gamma$--M and  $\Gamma$--K paths, respectively. These points are adjusted to achieve a good description of the top of the valence band and bottom of the conduction band by the TB model.  The convergence tolerance used was $10^{-11}$~eV. The bands considered were the valence and conduction bands around the Fermi level.

The first minimization consisted of finding the parameters corresponding to the SOC independent terms. Those were kept constant and a new minimization was performed to obtain the parameters corresponding to the SOC dependent terms. To improve the fit quality, a third minimization was carried out: parameters corresponding to the SOC dependent terms were kept fixed and those corresponding to the SOC independent terms were minimized. Lastly, the SOC independent terms were tested to verify whether they return band structures with Dirac points in the K and K$'$ points. The code for our TB model of Pt-based jacutingaites presented in this work is available at \cite{codeTB2024}.

\section{Results and discussion}
\subsection{Structural and electronic properties}\label{result_electronic}

The jacutingaite crystal presents hexagonal lattice and it also has a buckled honeycomb structure formed by $N$ elements, around a plane formed by the $M$ elements \cite{cabral2008platinum, vymazalova2012jacutingaite, de2020jacutingaite}, as shown in Figure \ref{structurefull}. For Pt-based jacutingaite, the lattice parameters and vertical buckling $N$--Pt--$N$ are presented in Table \ref{tabgap}. The obtained structural parameters are in good agreement with other theoretical predictions found in literature \cite{de2020jacutingaite}.

For Pt-based jacutingaite, in the absence of SOC, the hexagonal $N$--Pt--$N$ buckled structure gives rise to Dirac points at the K (and K$'$) points in momentum space. The projection of the energy bands near the Fermi level shows that the cone states are mostly composed by $N$ $2d$ transition metal 
(S$_{1/2}$ orbitals) hybridized with Pt $3e$ (5D$_{5/2}$ orbitals), which agrees with Reference \cite{de2020jacutingaite}. 

Some features of the Dirac cones in jacutingaite differ from those of graphene. In the latter, the Dirac cones exhibit linear dispersion around the Fermi level at each K and K$'$ points within the BZ. The introduction of SOC creates a gap of approximately 7 meV \cite{weeks2011engineering}. For Pt$_{2}NX_{3}$, the SOC opens a gap between 50 meV -- 154 meV at DFT-GGA-PBE level (Table \ref{tabgap}).
Already in this energy range is enable dissipationless charge transport at room temperature.
Moreover, there are regions of the energy bands along K(K$'$)--M, where the electron energy does not change significantly with respect to the momentum.

\begin{table}[h]
\caption{Lattice parameters $a$, vertical buckling $N$--Pt--$N$ $d_{z}$ and energy gap $E_g$ for Pt-based jacutingaite monolayers. \label{tabgap}}
\begin{indented}
\item[]
\begin{tabular}{L{1.6cm}|R{1.4cm} R{1.4cm} C{2cm}}
Pt$_{2}NX_{3}$ & ~~~$a$ (\AA) & ~~~$d_{z}$ (\AA) & $E_g$ (meV)\\ \hline \hline
Pt$_{2}$ZnS$_{3}$  & 7.173 &  3.000 & 150\\ 
Pt$_{2}$ZnSe$_{3}$ & 7.471 &  2.694 & 154\\ 
Pt$_{2}$ZnTe$_{3}$ & 7.924 &  2.259 &148\\ 
Pt$_{2}$CdS$_{3}$  & 7.187 &  3.689 & ~88\\ 
Pt$_{2}$CdSe$_{3}$ & 7.497 &  3.415 &~50 \\ 
Pt$_{2}$CdTe$_{3}$ & 7.935 &  3.019 &~73\\ 
Pt$_{2}$HgS$_{3}$  & 7.192 &  3.690 &118 \\ 
Pt$_{2}$HgSe$_{3}$ & 7.505 &  3.431 &150\\ 
\hline
\end{tabular}
\end{indented}
\end{table}

The honeycomb lattice and the sub-lattice symmetry of graphene is responsible for some of its unique electronic properties \cite{novoselov2004electric, katsnelson2007graphene, 
neto2009electronic}. On the other hand, the presence of quasi-flat behavior of electronic bands in M-K region may be exhibited by other structures, such as the kagome lattice. Both can present topological electronic states and robustness against perturbations. However, in the absence of a magnetic field, the topological phase transition occurs only when SOC is null and gaps closes at K (or K$'$) in the honeycomb lattice in the Kane-Mele model. The phase transitions in kagome lattice can be driven by the real NNN coupling and energy gap transitions with bands touching at other points in BZ different from K (and K$'$) \cite{beugeling2012topological}.
As the Pt $3e$ elements contribute significantly to the energy gap, the introduction of these two sublattices allows for the development of a more accurate TB model, which in turn may be used to study the properties of jacutingaite-based nanostructures. Some of those having been explored in the literature \cite{bafekry2020graphene, mauro2020multi, luo2021functionalization, rehman2022valley, zhu2023valley, de2023topological} but not extended to larger systems.

\begin{figure}[h]
\centering
\includegraphics[width=6cm]{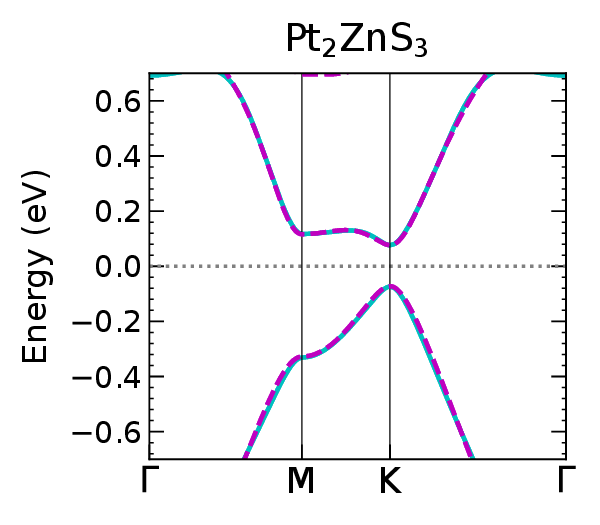}\includegraphics[width=6cm]{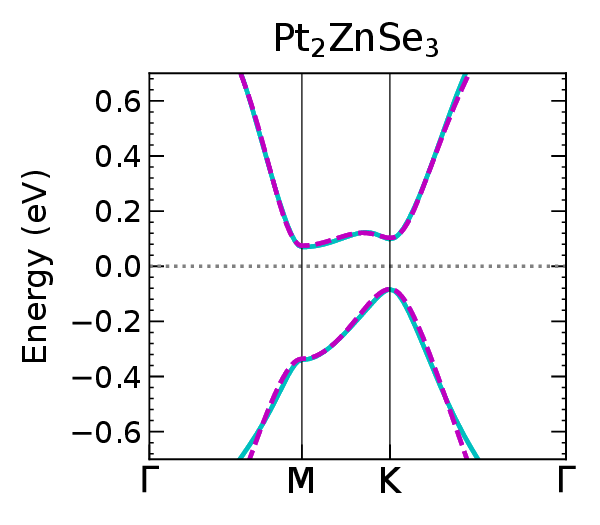}\includegraphics[width=6cm]{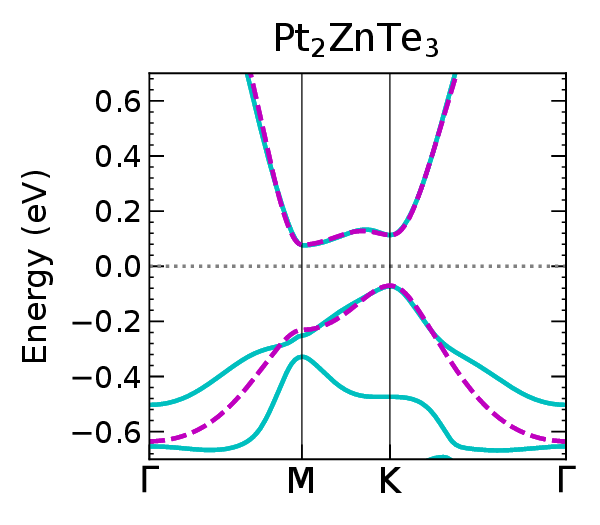}
\\
\includegraphics[width=6cm]{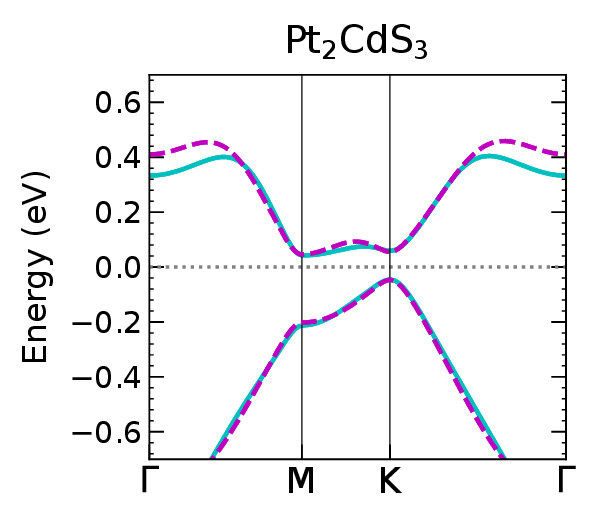}\includegraphics[width=6cm]{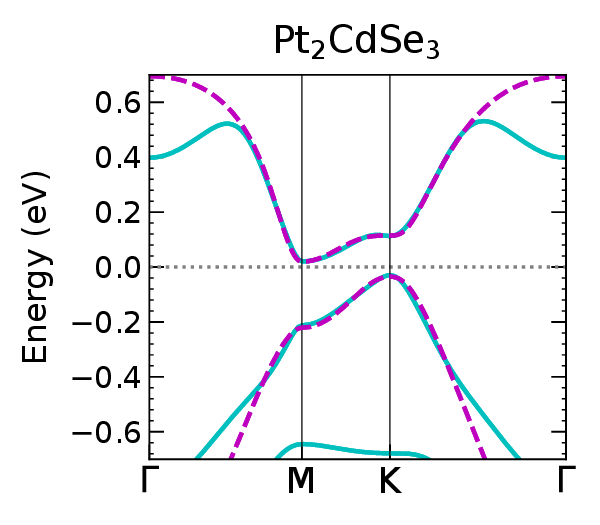}\includegraphics[width=6cm]{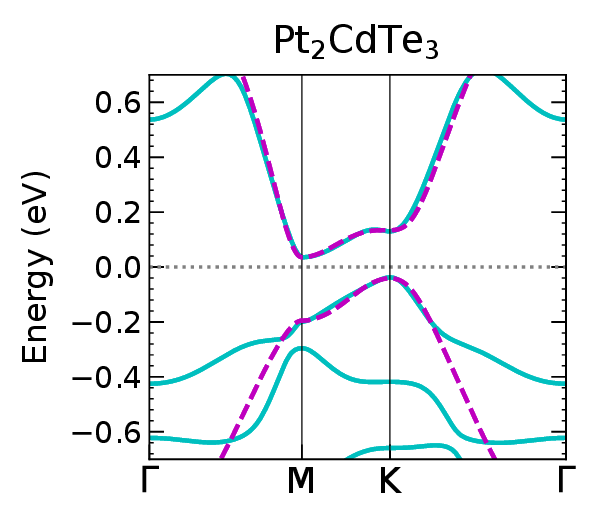}
\\
\includegraphics[width=6cm]{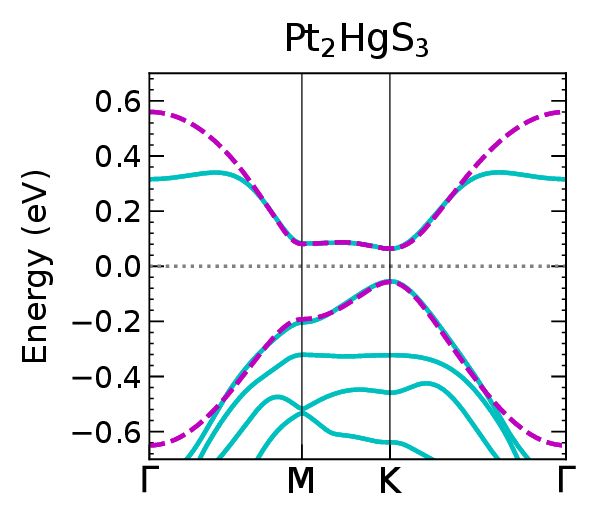}\includegraphics[width=6cm]{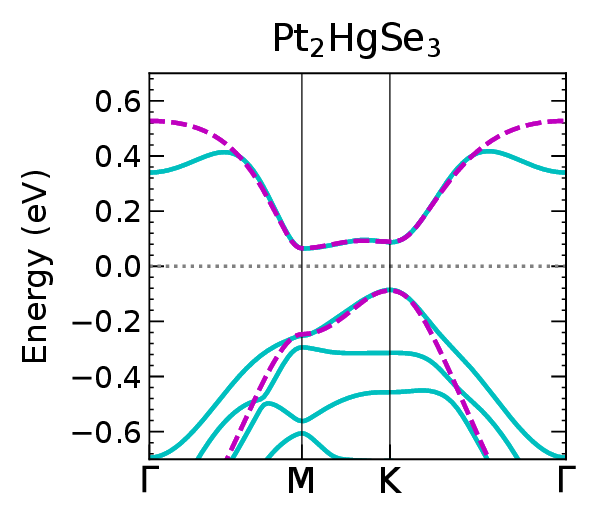}
\caption{Band structure for Pt-based jacutingaite for case with SOC. Fermi level is positioned to middle band gap. In blue line, energy bands obtained by DFT calculations. In pink dashed line energy bands obtained by TB model proposed. Note that in the M--K direction our TB model accurately describes the bands close to the Fermi level.} \label{strucband}
\end{figure}

Figure \ref{strucband} shows band structures for several Pt-based jacutingaite monolayers calculated by DFT-GGA (blue solid lines), as well as the results from the TB model (pink dashed lines). The TB results show very good agreement with the DFT bands within an energy interval of 0.6~eV to 1.0~eV and around the M--K path, which exhibits a  quasi-flat dispersion in the conduction bands. 

\subsection{Honeycomb-Kagome Tight-Binding model performance on jacuntigaite-like materials}

A reasonably accurate description of the electronic structure of the jacutingaite can be realized using an approach based on Maximally Localized Wannier Functions (MLWFs) \cite{marrazzo2018prediction}. The TB Hamiltonian obtained from the construction of MLWFs naturally involves a very large number of hopping parameters. For a structure such as that of jacutingaite, within the MLWFs formalism the number of hoppings terms is of the order of thousands. Thus, in order to study finite nanostructures and electronic transport properties, where a large real-space lattice is often required, the computational cost becomes very high. The model proposed in the present work is a TB Hamiltonian with an interaction radius reaching up to the $N$-NNN distance, and provides an accurate low-energy description with only twelve hopping parameters.

From the proposed model, and based on the band structures obtained via DFT, the twelve hopping parameters were defined for each monolayer structure and arranged in the Table \ref{tab_honeycomb}, 
\ref{tab_kagome} and 
\ref{tab_combination}, as shown in the Figure \ref{structurehopping} (a), (b) and (c), respectively, and for each term of the Equation \ref{hhk}.

\begin{table}[h]
\caption{Hopping amplitudes parameters for honeycomb lattice (in eV),  as shown in the Figure \ref{structurehopping} (a), where for  Zn, Cd and Hg the orbitals are 4S$_{1/2}$, 5S$_{1/2}$ and 6S$_{1/2}$, respectively.}\label{tab_honeycomb}
\begin{indented}
\item[]
\begin{tabular}{p{1.5cm}| R{1.4cm} R{1.4cm} R{1.4cm} R{1.4cm}}
   Pt$_{2}NX_{3}$& $\epsilon_{N}$ ~~~~~ & $t^{'}_{N-N}$~~~ & $t^{''}_{N-N}$~~~ & $\lambda^{''}_{N-N}$~~~ \\ \hline \hline
Pt$_{2}$ZnS$_{3}$  & -0.012& -0.359& 0.024& 0.025\\
Pt$_{2}$ZnSe$_{3}$ & 0.029& -0.406& 0.024& 0.032\\
Pt$_{2}$ZnTe$_{3}$ & 0.235& -0.376& -0.043& 0.022\\
Pt$_{2}$CdS$_{3}$  & 0.489& -0.420& 0.014& 0.021\\
Pt$_{2}$CdSe$_{3}$ & 0.023& -0.308& 0.036& 0.019\\
Pt$_{2}$CdTe$_{3}$ & 0.174& -0.362& 0.009& 0.012\\
Pt$_{2}$HgS$_{3}$  & 0.019& -0.217& 0.003& 0.017\\
Pt$_{2}$HgSe$_{3}$ & -0.039& -0.269& 0.032& 0.024\\
\hline
\end{tabular}
\end{indented}
\end{table}

\begin{table}[h]
\caption{Hopping amplitudes parameters for kagome lattice  (in eV),  as shown in the Figure \ref{structurehopping} (b), to orbital 5D$_{5/2}$ of the Pt elments. }\label{tab_kagome}
\begin{indented}
\item[]
\begin{tabular}{p{1.5cm}| R{1.4cm} R{1.4cm} R{1.4cm} R{1.4cm}}
   Pt$_{2}NX_{3}$& $\epsilon_{\textrm{Pt}}$ ~~~~ & $t^{'}_{\textrm{Pt-Pt}}$~~ & $t^{''}_{\textrm{Pt-Pt}}$~~ & $\lambda^{''}_{\textrm{Pt-Pt}}$~~ \\ \hline \hline
Pt$_{2}$ZnS$_{3}$ & 4.387& 0.398& -1.586& 0.045 \\
Pt$_{2}$ZnSe$_{3}$ & 4.548& 0.428& -1.635& 0.034\\
Pt$_{2}$ZnTe$_{3}$ & 5.291& 0.444& -1.957& 0.028\\
Pt$_{2}$CdS$_{3}$  & 4.100& 1.296& -0.375& 0.031\\
Pt$_{2}$CdSe$_{3}$ & 5.810& 0.716& -1.915& 0.344\\
Pt$_{2}$CdTe$_{3}$ & 7.502& 1.455& -1.820& 0.543\\
Pt$_{2}$HgS$_{3}$  & 8.487& 1.850& -1.870& 0.680\\
Pt$_{2}$HgSe$_{3}$ & 10.009& 0.986& -3.400 & 0.342\\
 \hline
\end{tabular}
\end{indented}
\end{table}

\begin{table}[h]
\caption{Hopping amplitudes parameters for combination honeycomb and kagome lattices  (in eV),  as shown in the  Figure \ref{structurehopping} (c), where for  Zn, Cd and Hg the orbitals are 4S$_{1/2}$, 5S$_{1/2}$ and 6S$_{1/2}$, respectively,  hybridized with 5D$_{5/2}$ of the Pt elements.}\label{tab_combination}
\begin{indented}
\item[]
\begin{tabular}{p{1.5cm}| R{1.4cm} R{1.4cm} R{1.4cm} R{1.4cm}}
   Pt$_{2}NX_{3}$ & $t^{'}_{\textrm{Pt}-N}$~~~ & $t^{''}_{\textrm{Pt}-N}$~~~ & $\lambda^{'}_{\textrm{Pt-Pt} }$~~~ & $\lambda^{''}_{\textrm{Pt}-N}$~~~\\ \hline \hline
Pt$_{2}$ZnS$_{3}$ & -0.145& -0.020& 0.019& 0.005 \\
Pt$_{2}$ZnSe$_{3}$ & -0.188& -0.029& 0.022& 0.012\\
Pt$_{2}$ZnTe$_{3}$ & -0.214& -0.054& 0.002& -0.008\\
Pt$_{2}$CdS$_{3}$  & -0.398& 0.202& 0.035& 0.012\\
Pt$_{2}$CdSe$_{3}$ & -0.199& -0.020& -0.407&0.014 \\
Pt$_{2}$CdTe$_{3}$ & -0.322& -0.006& -0.370& 0.023\\
Pt$_{2}$HgS$_{3}$  & -0.184& -0.074& 0.379& 0.000\\
Pt$_{2}$HgSe$_{3}$ & -0.256& -0.068& 0.037 &0.000 \\
 \hline
\end{tabular}
\end{indented}
\end{table}

The hopping terms in Table \ref{tab_honeycomb} are relatively similar for the different Pt-based jacutingaites, except for the $t^{''}_{\textrm{Zn}-\textrm{Zn}}$, which is a consequence of the more dispersive valence band of Pt$_{2}$ZnTe$_{3}$ as a way of compensating the SOC contribution from Te. 
The real NNN hopping terms are comparable to the SOC terms, which highlights its importance. In fact, if these terms are neglected, the results become qualitatively different from those obtained from first principles calculations. The negative value of $\lambda^{''}_{\textrm{Pt-Zn}}$ in Table \ref{tab_combination} is due to a greater energy difference between M and K(K$'$) points, being $E_{\textrm{M}}<E_{\textrm{K(K}'\textrm{)}}$ in the conduction band compared to Pt$_{2}$ZnSe$_{3}$.

For Pt$_{2}$CdSe$_{3} $, Pt$_{2}$CdTe$_{3}$, Pt$_{2}$HgS$_{3}$ and Pt$_{2}$HgSe$_{3}$ the  $\lambda^{'}_{\textrm{Pt-Pt}}$ and $\lambda^{''}_{\textrm{Pt-Pt}}$ is about an order of magnitude larger than $\lambda^{''}_ {N-N}$, due to a stronger hydridization between the orbitals 5D$_{5/2}$ and $h$S$_{1/2}$ for these monolayers.

In systems with strong SOC such as jacutingaite-like materials, the quantum numbers $|l j m_{j} m_{s}\rangle$ are strongly entangled and accurately describe the complexity  of the electronic states of these materials. In this work, we are not considering time reversal symmetry breaking, therefore, we do not evidence the $m_{j}$.

In Figure \ref{strucband} we show a comparison between the band
structures for each Pt-based jacutingaite monolayer calculated by DFT-GGA and that obtained from the TB model described in the previous section. The TB model for interpenetrating honeycomb and kagome lattices shows good agreement with the DFT-GGA results and is quite accurate along the K(K$'$)--M direction. We emphasize that this level of accuracy is missing in previous studies, where simpler TB Hamiltonians based on the Kane-Mele model were employed.

\begin{figure}[h]
\centering
\includegraphics[width=9cm]{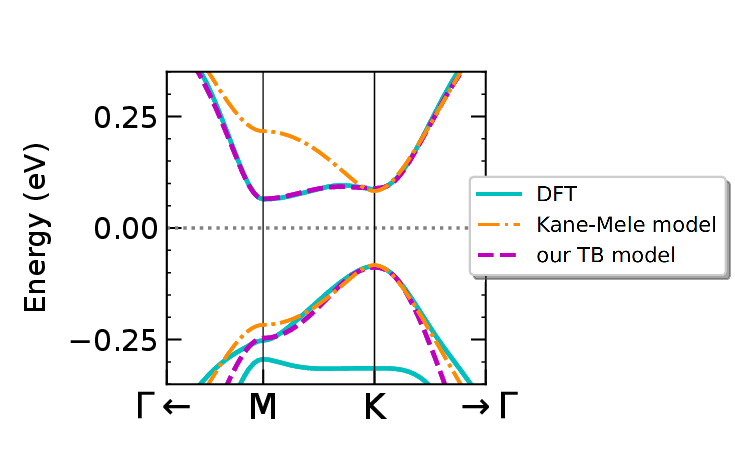}
\caption{
Comparison between the energy band structures of the proposed TB model (pink dashed line) and the Kane-Mele model (orange dash-dot line) for Pt$_2$HgSe$_3$, with the latter adjusted to better agreement with the energy bands obtained from DFT calculations (blue line). Notably, around the M point, the energy dispersion behavior of the conduction band is not accurately described by the Kane-Mele model, in contrast to the proposed TB model which effectively captures this behavior.
} \label{ourmodel_km}
\end{figure}

It is worth pointing out that quasi-flat bands are not seen in the Kane-Mele model and the states along K(K$'$)--M present prominent features of the kagome lattice band, as illustrated in Figure \ref{ourmodel_km} for the Pt$_2$HgSe$_3$.  The same qualitative behavior is observed for the other considered materials.

\begin{figure}[h]
\centering
\includegraphics[width=8.4cm]{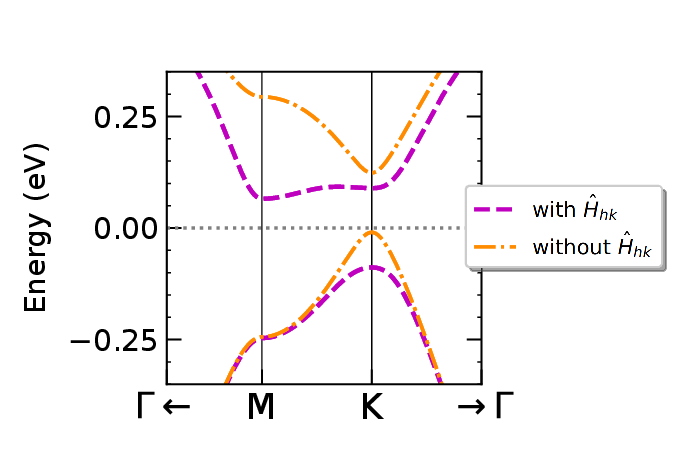}
\caption{
Comparison between the energy band structures of the proposed TB model without (salmon dash-dot line) and with (pink dashed line) $\hat{H}_{\mathrm{hk}}$ Hamiltonian for Pt$_2$HgSe$_3$. Notably, the inclusion  $\hat{H}_{\mathrm{hk}}$ term contributes strongly to the description of the conduction band dispersion characteristic of jacutingaite.}
 \label{ourmodel_coupling}
\end{figure}

The $\hat{H}_{\mathrm{hk}}$ term, which couples the honeycomb and kagome sub-lattices, as given by Equation \ref{keyEq}, is responsible for the characteristic shape of the conduction band of jacutingaite, specifically the band flattening along M--K(K$'$) direction. In Figure \ref{ourmodel_coupling} one can visualize, for the case of Pt$_2$HgSe$_3$, the energy dispersion both with and without the inclusion of such term in  Equation \ref{hhk}. The relevance of considering this coupling when dealing with Pt-based jacutingaite using tight-binding models thus becomes evident.

For Pt-based jacutingaites monolayers, in the absence of SOC, a calculation of the hybrid Wannier center (HWC) from the present TB model, one can observe that they are non-topological systems (see SM, Figure 2 \cite{materialsupplementary}). The SOC thus opens the topological energy gap and that information is also captured by this model (see SM, Figure 3 \cite{materialsupplementary}).

\subsection{Jacuntigaite nanoribbons}

The calculation of topological invariants allows determining the topological character of a material based on its bulk structure \cite{vanderbilt2018berry}, which is very useful for DFT simulation of crystalline systems. 
Another approach is to explore topological properties in materials is to consider finite size systems and investigate their edge or surface states. The combination of DFT simulations with the second strategy often implies high computational cost, which can be avoided by the use of a DFT derived TB model.

Edge states in topological 2D systems are topologically protected and robust against local perturbations \cite{hasan2010colloquium, bansil2016colloquium}. In case of Pt-based jacutingaite monolayers, which are QSH semiconductors, the edge states are protected by time-reversal symmetry and against backscattering processes, presenting robustness against non-magnetic perturbations \cite{kandrai2020signature, marrazzo2018prediction}.
Understanding and manipulating these edge states not only defines the stability of materials, but also diversifies the possibilities for technological applications \cite{bhardwaj2020topological, bernevig2006quantum, weeks2011engineering, beugeling2012topological, bafekry2020graphene, rehman2022jacutingaite}.

\begin{figure}[h]
\includegraphics[width=18cm]{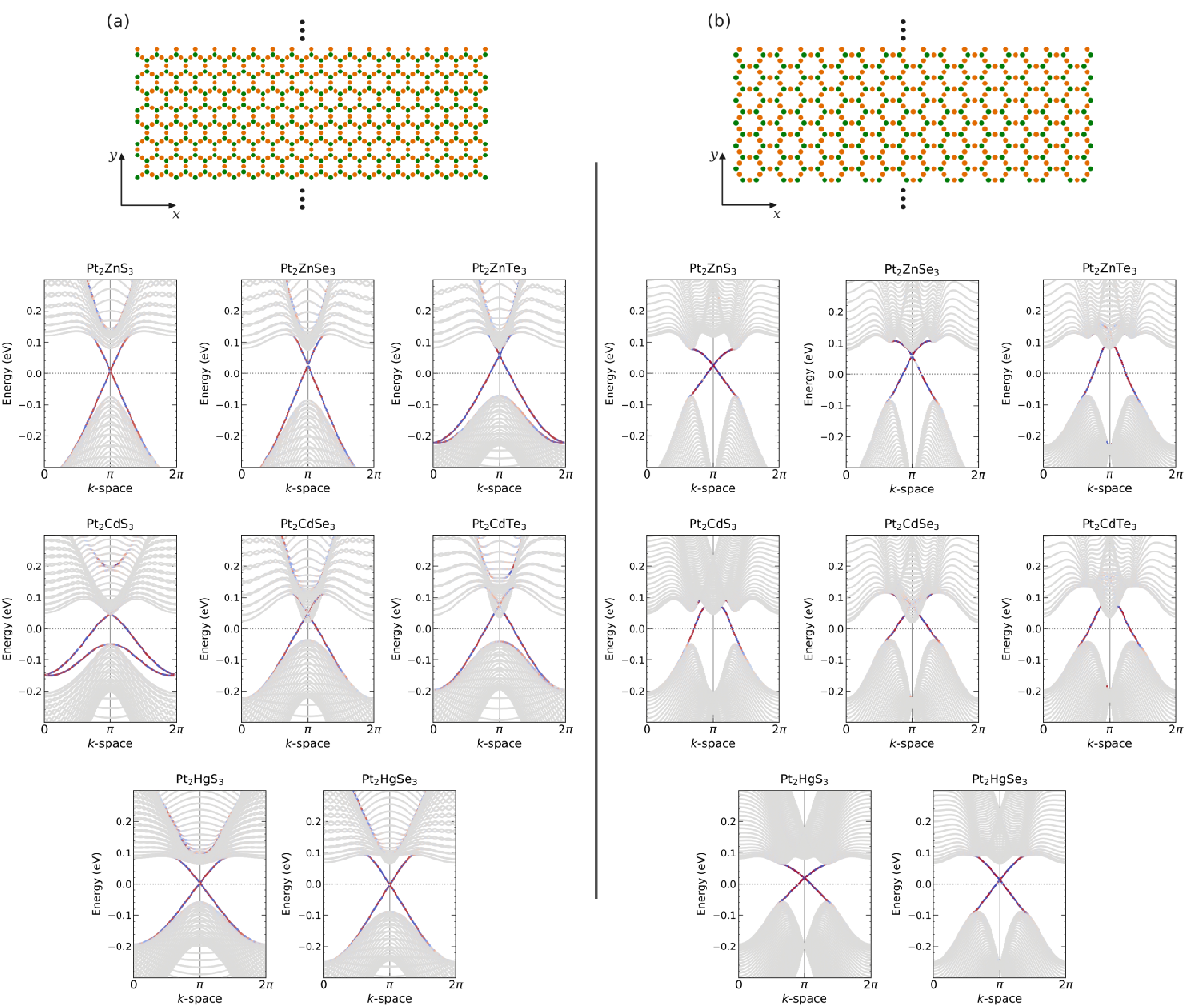}
\caption{Nanoribbons (a) armchair and (b) zigzag type with symmetrical edge.
Band structure, where light gray are the bulk bands, red and blue are the edge states on the two opposite edges of nanoribbon (c) armchair and (d) zigzag type with width $30a$ and $30\sqrt{3}a$, respectively, being $a$ lattice parameters present in Table \ref{tabgap}. There are two pairs of edge states equal in energy and in momentum.} \label{nanoribbons_ss}
\end{figure}

We consider Pt-based jacutingaites nanoribbons with symmetric and asymmetric opposite edges. Figure \ref{nanoribbons_ss} (a) and (b) shows nanoribbons with symmetric armchair and zigzag opposite edges. There is a pair of topological states of opposite spin at each edge connecting the valence and conduction bulk states, as we can visualize in their respective band structure in  Figure \ref{nanoribbons_ss}. There are no significant differences in the edge states considering a symmetric or asymmetric configuration of the edge for all cases of Pt-based jacutingaites. The edges have the same geometry, with just a difference in translation in relation to each other and there are no overlap significant between edge states to change its energy dispersion. Since a nanoribbon is wide enough for the interaction between the edges not to destroy the topological character, the topological edge states are characterized predominantly by SOC.

The inclusion of NNN hopping terms can lead to the emergence of edge states, which has been a focus of recent studies \cite{downing2024unconventional}. The presence of these states, however, depends on the edge geometry, presence of defects, vacancies and external perturbations. In Figure \ref{nanoribbons_ss} (a) it is possible to observe the existence of edge states within the energy spectrum of the conduction bands. These states emerge as a consequence of the mapping of asymmetric interactions (Figure \ref{structurehopping} (b) and (c)).

\begin{figure}[h]
\includegraphics[width=8cm]{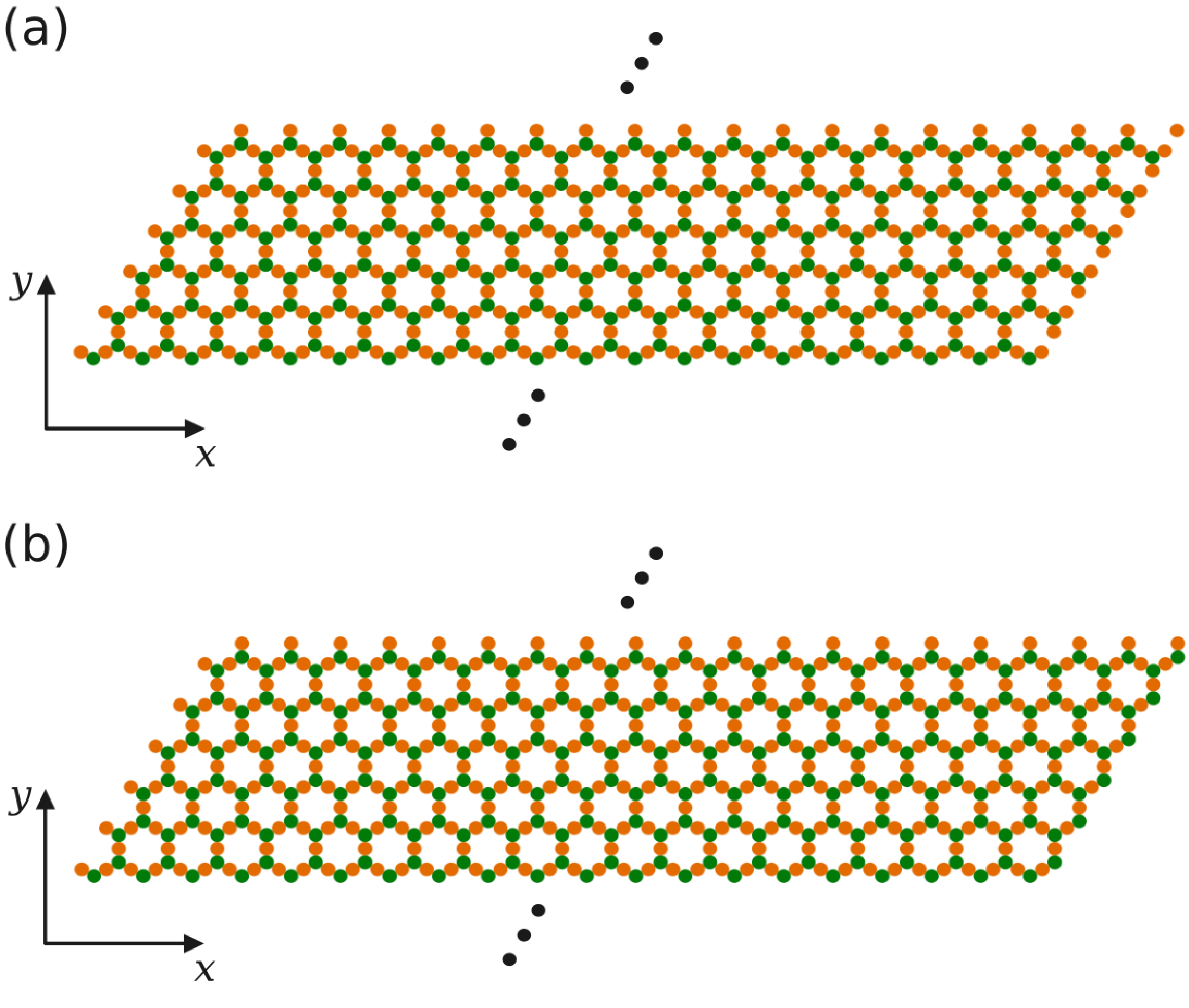}
\includegraphics[width=8cm]{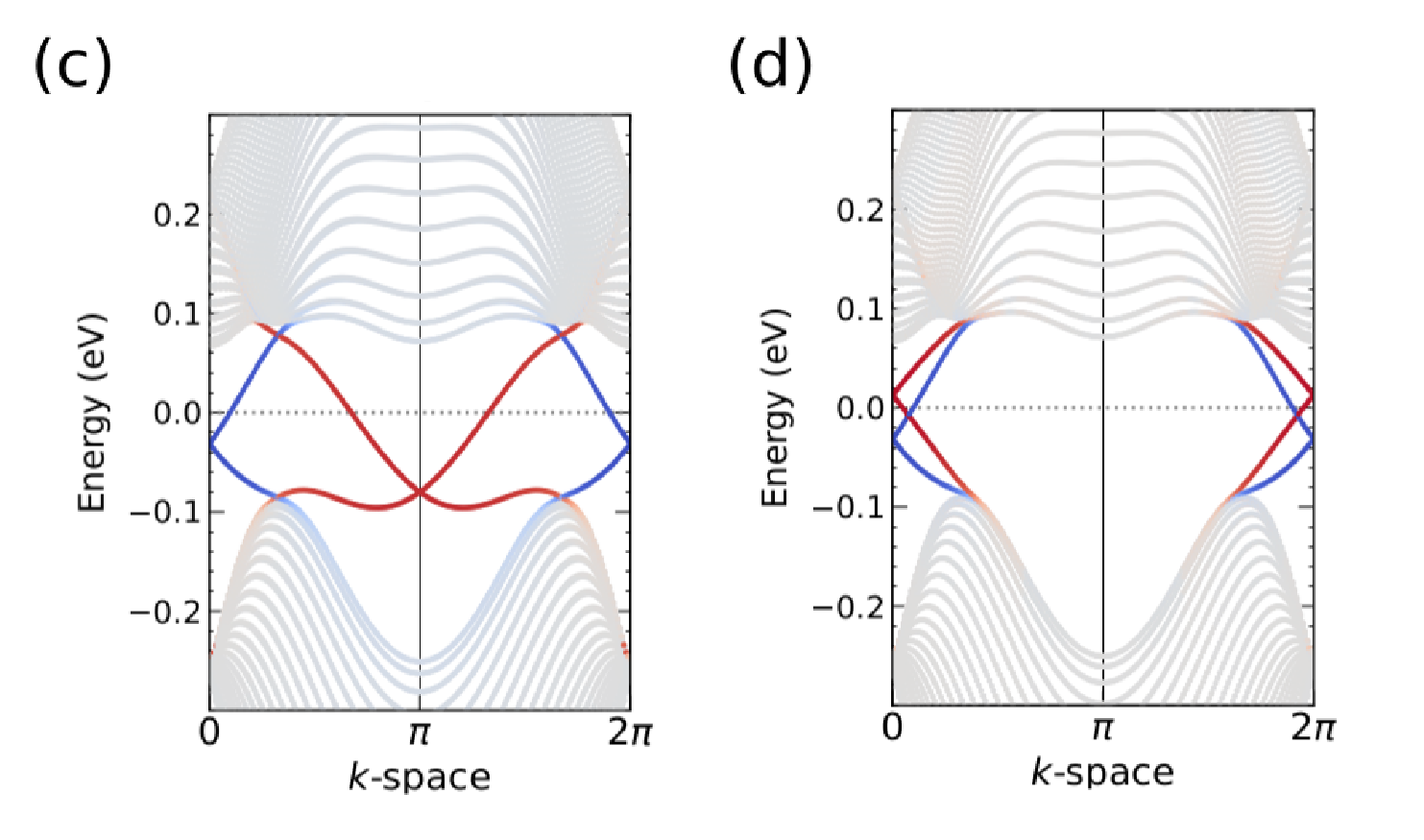}
\caption{Nanoribbons zigzag type with different non-symmetric edge (a) and (b).
Band structure (c) and (d), where light gray are the bulk bands, red and blue are the edge states on the two opposite edges of nanoribbons represent in (a) and (b), respectively with width 22.5 nm for the case of Pt$_{2}$HgSe$_{3}$. There are two pairs of edge states different in energy and momentum in (a, c) and different in energy only in (b, d).} \label{nanoribbons_as}
\end{figure}

Two Pt$_{2}$HgSe$_{3}$ nanoribbons with non-symmetric edges are represented in Figure \ref{nanoribbons_as} (a) and (b).
For the case Pt$_{2}$HgSe$_{3}$ nanoribbon represents in Figure \ref{nanoribbons_as} (a) there is a small energy shift and a large momentum shift between the topological states of opposite edges, as shown in  Figure \ref{nanoribbons_as} (c). For the case of the nanoribbon represents in  Figure \ref{nanoribbons_as} (b) there is only a small energy shift between the topological states of opposite edges (see Figure \ref{nanoribbons_as} (d)). For other Pt-based jacutingaite nanoribbons see SM,  Figure 4 and 5. 

\begin{figure}[h]
\includegraphics[width=7.8cm]{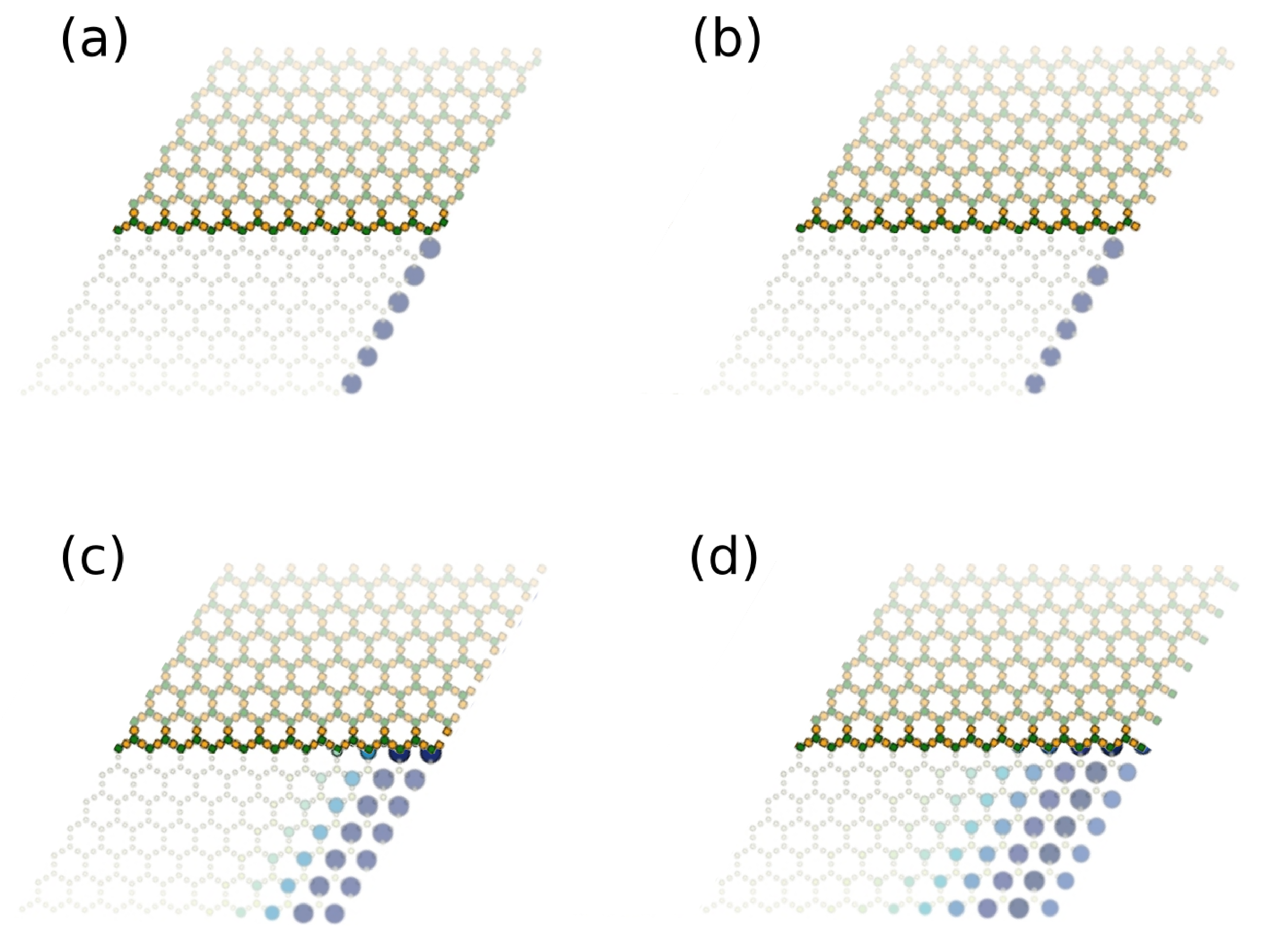}
\caption{Possible geometric configurations of Pt-based jacutingaite edges, based on our TB model, and the representation of the proportional penetration length of the edge states for Pt$_{2}$HgSe$_{3}$ nanoribbons for the respective terminations. Green and orange spheres represent $N$ and Pt $3e$ elements, respectively. In (a) and (b), the states tend to be often localized, while for (c) and (d) the edge states tend to be significantly more delocalized.} \label{edges}
\end{figure}

Depending on the geometric configuration of the edges, the topological edge states can be highly localized at the edges or be more delocalized. In Figure \ref{edges}, there is a representation of the proportional penetration length of the edge states for Pt$_{2}$HgSe$_{3}$ nanoribbons with some possible terminations. For the configurations represented in Figure \ref{edges} (a) and (b), the states tend to be often localized, confined to a distance of up to $a$, for both cases, with the value of $a$ given in Table \ref{tabgap}. For the configurations illustrated in  Figure \ref{edges} (c) and (d), the edge states tend to be significantly more delocalized, with a penetration length of up to 7$a$ and 11$a$, respectively. A nanoribbon of width 85$a$ was considered for these terminations.

These results show that the corresponding dispersion bands and penetration length of the edge states to the topological edge states of Pt-based jacutingaite finite nanostructures are sensitive to the geometry of the edge. Additionally, these results demonstrate that the TB model can go beyond the computational limitations associated with DFT computational costs. The presented approach opens an avenue to explore large-scale structures in a computationally resource-efficient approach.

\section{Conclusion}

In this study, we have successfully developed a comprehensive TB model for Pt-based jacutingaite monolayers, meticulously capturing both the conduction and valence band dispersions near the SOC energy gaps. Our model uniquely integrates the linear dispersion characteristics from the K(K$'$)--$\Gamma$  points and the downward band from K(K$'$) to M, significantly enhancing accuracy by incorporating Pt 3e elements. These elements, in conjunction with the $N$  metals, form an intricate lattice that is a hybrid of honeycomb and kagome sub-lattices, further enriched by real next-nearest-neighbor (NNN) hopping terms.
Considering this model, we were able to reproduce the quasi-flat bands, not seen in the Kane-Mele model, and the states along K(K$'$)–M, which present prominent features of the kagome lattice band.

The implications of our model are profound, offering a new lens for analysing of the system to obtain more reliable TB Hamiltonians. Our TB model, by accurately describing low-energy regions, enables an in-depth exploration properties of mesoscopic systems, as demonstrated by our analysis of various nanoribbon systems and its topological properties. Furthermore, experimental techniques have been used to explore the robustness of topological states under real conditions. Advances in scanning tunneling microscopy techniques can detect the simultaneity of bulk gaps and metallic edge states \cite{kandrai2020signature, yin2021probing, weber20242024}. The observation of hybridization effects is enable by high-resolution angle-resolved photoemission measurements at different photon energies \cite{cucchi2020bulk, lv2019angle, weber20242024}. Both the techniques have already been employed in the studies of the Pt$_{2}$HgSe$_{3}$, to demonstrate the dual-topological nature in its bulk form \cite{cucchi2020bulk} and the existence and robustness of QSH states in its 2D form \cite{kandrai2020signature}. Theoretical models help in the interpretation and understanding of experimental results, and our model is very promising in playing this role in the future experimental study of Pt-based jacuntigaites.

The potential applications of this model are vast. Our findings contribute not only  to a deeper understanding of Pt-based jacutingaite monolayers but also pave the way for future research in this exciting field. Further studies could explore the applicability of our model to other 2D materials or more complicated systems, building on the materials studied in this paper, potentially uncovering new phenomena and guiding the design of next-generation electronic devices.

\section*{Acknowledgment}

We acknowledge prof. Cristiane Morais Smith, from the Institute of Theoretical Physics at Utrecht University, for the insightful discussions that inspired this work.

This work was financially supported by the Brazilian Council for Research (CNPq) Grants 310422/2019-1

\section*{References}

\end{document}